\documentclass[twocolumn]{aastex631}

\usepackage{graphicx}	
\usepackage{amsmath}	
\usepackage{CJKutf8}
\usepackage{multirow}
\usepackage{units}
\usepackage{hyperref}

\shorttitle{The spin-up line of MSPs}
\shortauthors{Liu et al.}
\graphicspath{{./}{figures/}}

\begin{document}

\begin{CJK*}{UTF8}{gbsn}

\title{Observational evidence for a spin-up line in the P-Pdot diagram of millisecond pulsars}


\correspondingauthor{Xing-Jiang Zhu}
\email{zhuxj@bnu.edu.cn}

\author[0000-0002-2187-4087]{Xiao-Jin Liu}
\author[0000-0002-3309-415X]{Zhi-Qiang You}
\affiliation{Advanced Institute of Natural Sciences, Beijing Normal University, Zhuhai 519087, China}
\affiliation{ Department of Astronomy, Beijing Normal University, Beijing 100875, China}

\author[0000-0001-7049-6468]{Xing-Jiang Zhu}
\affiliation{Advanced Institute of Natural Sciences, Beijing Normal University, Zhuhai 519087, China}



\begin{abstract}
It is believed that millisecond pulsars attain their fast spins by accreting matter and angular momentum from companion stars.
Theoretical modelling of the accretion process suggests a spin-up line in the period-period derivative ($P$-$\dot{P}$) diagram of millisecond pulsars, which plays an important role in population studies of radio millisecond pulsars and accreting neutron stars in X-ray binaries.
Here we present observational evidence for such a spin-up line using a sample of 143 radio pulsars with $P<\unit[30]{ms}$.
We also find that PSRs~J1823$-$3021A and J1824$-$2452A, located near the classic spin-up line, are consistent with the broad population of millisecond pulsars.
Finally, we show that our approach of Bayesian inference can probe accretion physics, allowing constraints to be placed on the accretion rate and the disk–magnetosphere interaction.

\end{abstract}


\section{Introduction} \label{sec:intro}

It has long been theorized that pulsars in binary systems can get spun up through accretion of matter from their companions in a so-called recycling process \citep[e.g.,][]{BhatVanden91,lor08}.
The first binary pulsar, B1913+16, was observed with a spin period of $\unit[59]{ms}$ in a tight eccentric orbit \citep{HulseTaylor75}.
It was soon realized that its progenitor was likely to be a high-mass X-ray binary \citep{Flannery75,DeLoore75}, with the pulsar experiencing an episode of mass accretion from an evolved companion star \citep{SmarrBlandford76}.
The discovery of the first millisecond pulsar (MSP) B1937+21 \citep{Backer82msp} reinforced the pulsar binary-recycling model \citep{acr+82,RadhakSrini82}, even though it appeared to be an isolated pulsar.
The recycling theory is strongly supported by observations of accreting millisecond X-ray pulsars \citep{Wijnands98,pw21} and transitional MSPs that switch between rotation-powered radio and accretion-powered X-ray states \citep{asr+09,Jaodand2016J1023,Papitto13Nat}, and by the fact that a majority of MSPs are found in binary systems.

Among the currently known 3300+ pulsars, MSPs represent a distinct population with spin periods $P \lesssim \unit[30]{ms}$ and slow spin-down rates $\dot{P}\lesssim 10^{-18}$, clustered in the lower-left corner of the pulsar $P$-$\dot{P}$ diagram; see the latest version of the ATNF Pulsar Catalogue\footnote{\url{https://www.atnf.csiro.au/research/pulsar/psrcat/}} \citep{mht+05}.
The existence of a spin-up line in the $P$-$\dot{P}$ diagram can be demonstrated by considering that: at the end of the recycling process, 1) the pulsar reaches an equilibrium spin period approximated by the Kepler orbital period of matter at the pulsar magnetospheric boundary \citep{pr72, do73, vdh77, gl79}, and 2) the magnetic dipole field strength at the surface is estimated as $B\propto \sqrt{P\dot{P}}$ \citep{spi06}.
The expression of the spin-up line takes the form \citep[e.g.,][]{GhoshLamb92,tlk12}
\begin{equation}
 	\label{eqn:spin_up_model}
	\dot{P} = A P^{4/3}
\end{equation}
where $A$ is a coefficient depending on the accretion rate, the magnetic inclination angle and accretion disk-magnetosphere interaction, among other factors.
Extending the magnetic dipole to multipoles leads to a more general form of the spin-up line \citep{aro93}
\begin{equation}
 	\label{eqn:spin_up_l}
	\dot{P} = \mathcal{A} (P/{\rm{s}})^{2-(2l/3)}\ ,
\end{equation}
where $l$ is the order of the magnetic multipole.

The spin-up line has been used as a powerful tool in various studies of binary and millisecond pulsars.
For example, it provides a more reliable age estimate than the characteristic spin-down age for recycled pulsars \citep[e.g.,][]{acw99,Kiziltan10msp}.
This is essential in deriving the effective lifetimes of Galactic double neutron star systems and their merger rate \citep{acw99,Burgay03dpsr,Pol19DNSrate}.
The spin-up line also acts as a useful link between millisecond radio pulsars and accreting neutron stars in X-ray binaries, facilitating the study of different pulsar sub-populations and their connection \citep[e.g.,][]{Kiziltan09,Ho14}.

The location of the spin-up line plays a critical role in understanding the characteristics of MSPs, such as their spin evolution and the amount of accreted masses \citep{tlk12}.
In \cite{acw99}, a fiducial value of $(1.1 \pm 0.5)\times10^{-15}$~s$^{-4/3}$ was given for the Eddington-limited accretion rate. However, uncertainties are likely to be more significant due to the lack of understanding of the accretion process and magnetic geometry \citep{tlk12, vf14}. Given the fiducial value and allowing small uncertainties, the spin-up line seems to be consistent with the vast majority of MSPs discovered up to date, but faces a challenge to explain two pulsars, J1823$-$3021A \citep{bbl+94, faa+11} and J1824$-$2452A \citep{fbt+88, jgk+13}, for which the observed spin-down rates are above $10^{-18}$, much higher than those of other MSPs.

In this Letter, we first search for observational evidence of a spin-up line in the $P$-$\dot{P}$ diagram of MSPs.
We collate a sample of radio MSPs with reliable measurements of intrinsic $\dot{P}$ and use them to perform Bayesian model selection of the joint $P$-$\dot{P}$ distribution with and without a spin-up cutoff, i.e., whether or not all MSPs are located to the lower-right side of a line consistent with the $\dot{P}\sim P^{4/3}$ relation.
We find that the model with a spin-up line is significantly favoured by the data.
We further demonstrate that the location of the spin-up line inferred from our analysis allows constraints to be placed on the accretion physics.

The remainder of this paper is structured as follows.
We introduce the pulsar data in Section~\ref{sec:pulsar_data}.
We describe the population model and Bayesian inference framework in Section~\ref{Bayesian_framework}.
Our results are presented in Section~\ref{paper_results}.
Finally, we discuss implications of our results and provide concluding remarks in Section~\ref{sec:conclu}.

\section{The pulsar data} 
\label{sec:pulsar_data}

In this work, we consider radio pulsars with $P< \unit[30]{ms}$; since there is not an exact spin period cut for the definition of an MSP, we discuss the analysis results if we apply the cut at $\unit[10]{ms}$ in the Appendix.
We focus on pulsars with measured spin-down rates $\dot{P}_{\rm obs}$ $\ge 10^{-20}$~s~s$^{-1}$ in the ATNF Pulsar Catalogue, i.e., selecting only recycled MSPs in the upper part of the $P$-$\dot{P}$ diagram since they provide the most constraining power for a spin-up line.
It is well known that $\dot{P}_{\rm obs}$ is usually contaminated by kinematic or dynamic effects.
Therefore, we further check and pick out the MSPs that have well-constrained intrinsic spin-down rates ($\dot{P}_{\rm int}$).

The $\dot{P}_{\rm int}$ and errors used in the analysis are obtained as follows.
For the pulsars in the Galactic field, both the contribution from Shklovskii effect \citep{shk70} and Galactic potential \citep{dt91, lbs18} are corrected.
Among these pulsars, 36 have reported $\dot{P}_{\rm int}$ in the literature thus the reported values are used directly. 
For the correction of remaining field pulsars, we use the proper motion in the ATNF Pulsar Catalogue when available, otherwise, assume a 2D transverse velocity of $87\pm13$~km~s$^{-1}$ \citep{HLL+05} 
In our calculations, we use the best-estimate distances from the ATNF Pulsar Catalogue, which by default are derived from measured dispersion measures in combination with the Galactic electron density model of \citet{ymw17}; for the distance estimate we assume an uncertainty of 30 percent. Note that the assumed distance uncertainty is somewhat arbitrary as there is no uncertainty given for the distance estimate in the ATNF catalogue. However, we have verified that our results are insensitive to this quantity because the inference is dominated by Galactic-field pulsars close to the spin-up line; for these pulsars their $\dot{P}_{\rm int}$ are very well constrained.

For pulsars in globular clusters, if the pulsar is in a binary system and has a measured orbital period derivative ($\dot{P}_{\rm b}$), $\dot{P}_{\rm obs}$ are corrected using $\dot{P}_{\rm b}$, assuming negligible gravitational radiation. 
Otherwise, the contribution from the cluster potential is difficult to measure but usually much larger than the contribution from Shklovskii effect and Galactic potential \citep{phi93}. For these pulsars, we neglect the latter two effects and use the maximum contribution from the globular cluster potential to set a bound on $\dot{P}_{\rm int}$ \citep[e.g.][]{lfr+12}. In this case, we only include the pulsars with a positive lower bound (i.e., $\dot{P}_{\rm int}>0$ after correction).

\begin{figure}[htp]
	\centering
	\includegraphics[width=1\columnwidth]{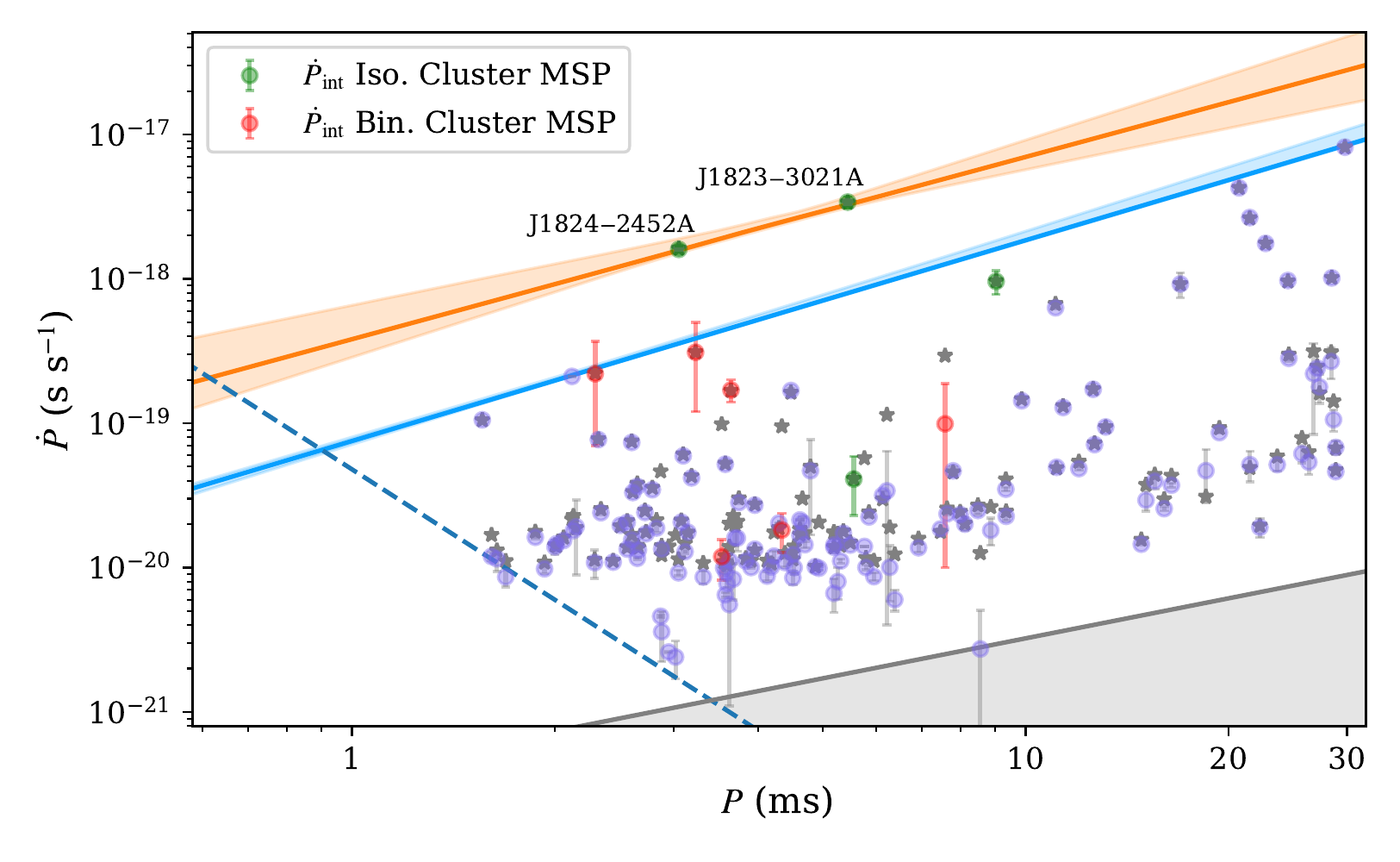}		
	\caption{The $P$-$\dot{P}$ diagram of the MSPs used in the analysis. The observed $\dot{P}$ is shown by gray stars, while the corrected $\dot{P}$ is shown by (filled) circles with error bars. The isolated and binary cluster MSPs are colored in green and red, respectively, while the Galactic-field pulsars are in purple. The orange and light blue bands are the 90\% credible intervals of the spin-up line if we include or exclude two ``anomalous" pulsars (J1823$-$3021A and J1824$-$2452A), respectively. The dashed line represents a constraint set by gravitational radiation due to a minimum ellipticity $\epsilon=10^{-9}$ \citep{wph+18}; plotted assuming $I=10^{38}$~kg~m$^2$. 
	The gray line with shaded region indicates the pulsar death valley \citep{rr94}. The P-Pdot data to reproduce this plot are available at \url{https://github.com/zhuxj1/MSPspinUp}.
	}
	\label{fig:PPdot}
\end{figure}

In total, we obtain 143 pulsars, where 133 pulsars are in the Galactic field and the remaining 10 are from globular clusters.
Among the globular cluster pulsars, three pulsars are in binary and have measurable $\dot{P}_{\rm b}$ while the remaining seven are isolated or have no measurable $\dot{P}_{\rm b}$.
Figure~\ref{fig:PPdot} shows the $P$-$\dot{P}$ diagram of 143 pulsars. In the plot, $\dot{P}_{\rm obs}$ is shown by gray stars while the intrinsic $\dot{P}$ is shown by filled circles. The error bar represents the 1-$\sigma$ uncertainty of $\dot{P}_{\rm int}$ for pulsars with $\dot{P}_{\rm b}$ in the Galactic field and for binary pulsars in globular clusters. For other pulsars in globular clusters, the error bar indicates the range of possible $\dot{P}_{\rm int}$ after considering the maximum contribution from the cluster potential.
Note that $\dot{P}_{\rm int}$ are constrained relatively well for the cluster MSPs shown in Figure \ref{fig:PPdot}; we provide further details about these pulsars in the Appendix.

As mentioned in the Introduction, the spin-down rates of PSRs J1823$-$3021A and J1824$-$2452A are much larger than those of the remaining pulsars in the data set. These two pulsars could belong to a different sub-class. We therefore consider two cases in our analysis depending on including these two pulsars or not.

\section{The Method} \label{Bayesian_framework}

\subsection{The spin-up line model}
\label{sec:model}

We first search for observational evidence of a spin-up line given by the general form of Equation (\ref{eqn:spin_up_l}).
We adopt a log-uniform prior for $\mathcal{A}$ in the range of $[10^{-17}, 10^{-12}]$.
Note that for a parameter that spans orders of magnitude, it is customary in Bayesian inference to adopt a log-uniform prior\footnote{We find that our results are not affected by the choice of prior because the posteriors are dominated by the likelihood.}.
For the magnetic multipole parameter, $l=0,1,2,3$ corresponds to pure magnetic monopole, dipole, quadrupole, hexapole, respectively.
Simply for the convenience of analysis, we adopt a uniform prior for $l \in [0, 3]$, or equivalently a power exponent in $[0, 2]$ for Equation~(\ref{eqn:spin_up_l}).
The competing hypothesis here is that there is no spin-up line cutoff in the joint $P$-$\dot{P}$ distribution.
Since in practice we have to set a range for the joint $P$-$\dot{P}$ distribution anyway, the competing hypothesis is taken to be one that has a horizontal cut-off line (at an unknown maximum $\dot{P}$) in the $P$-$\dot{P}$ diagram.

Next we investigate the usefulness of the spin-up line in constraining accretion physics by adopting the model of \cite{tlk12}, where the coefficient $A$ of Equation (\ref{eqn:spin_up_model}) is given by
\begin{equation} 
	\label{eqn:spin-up-param}
	A = \frac{2^{1/6}G^{5/3}}{\pi^{1/3}c^3}\frac{\dot{M}M^{5/3}}{I}(1+\sin^2\alpha)\phi^{-7/2}\omega_{\rm c}^{7/3},
\end{equation}
where $G$ is the gravitational constant, $c$ is the speed of light, $\dot{M}$ is accretion rate, $M$ is pulsar mass, $I$ is moment of inertia, $\alpha$ is magnetic inclination angle, $\phi$ is a parameter relating the magnetospheric boundary $r_{\rm{mag}}$ to the Alfvén radius $r_{\rm{A}}$ so that $\phi=r_{\rm{mag}}/ r_{\rm{A}}$ \citep{wang96, ga19}, and $\omega_{\rm c}$ is the critical fastness parameter \citep{gl79, ga19}. Note that Equation (\ref{eqn:spin-up-param}) takes into account the plasma contribution to the spin-down torque \citep{spi06}, thus differs from the result of assuming vacuum dipole magnetic field. 

For the sake of convenience, we scale $\dot{M}$ by Eddington accretion rate ($\dot{M}_{\rm Edd}$) via $\dot{M} =r\dot{M}_{\rm Edd}$. $\dot{M}_{\rm Edd}$ has a weak dependence on the fraction of hydrogen mass ($X$):  $\dot{M}_{\rm Edd} \propto 1/(1+X)$, where $X\le0.2$ for the late phase of accretion \citep{tlk12}. Without losing much precision, we used $X=0.2$ throughout this paper and consider $r$ in the range of [0.01, 3] to allow for possible super-Eddington accretion \citep[e.g.,][]{faa+11}. For the moment of inertia, we use the following relation \citep{ls05}: 
\begin{equation} 
	\label{eqn:inertia}
	 I = 0.237 MR^2\Bigg[1+4.2\frac{M}{M_\odot}\frac{\rm km}{R} + 90\Bigg(\frac{M}{M_\odot}\frac{\rm km}{R} \Bigg)^4 \Bigg], 
\end{equation}
where $R$ is the pulsar radius. This relation is valid for $M/R\ge0.07$~M$_\odot$~km$^{-1}$ and suitable for a wide range of neutron star equations of state.  

In an attempt to probe accretion physics, we fix the spin-up line exponent ($\dot{P} \sim P^{4/3}$) and derive constraints on parameters in Equation (\ref{eqn:spin-up-param}).
We use a log-uniform prior for the accretion rate parameter $r$ and uniform priors for $M, R, \alpha, \phi$ and $\omega_{\rm c}$. The prior ranges of the parameters are listed in Table~\ref{tab:spinup_param}.
Note that, as pointed out by \citet{tlk12}, the spin-up line is not uniquely defined; instead it depends on pulsar-specific parameters such as the mass and radius.
Therefore, the spin-up line considered here is an envelope that encapsulates the MSP population in question.

\subsection{Bayesian inference}

Assuming that the distribution of pulsars in the $P$-$\dot{P}$ diagram can be described by a model consisting of a set of hyper-parameters $\Lambda$, then within the Bayesian inference framework, we have 
\begin{equation}
    p(\Lambda|d) = \frac{\mathcal{L}(d|\Lambda)\pi(\Lambda)}{Z_d},
\end{equation}
where $\mathcal{L}(d|\Lambda)$ is the total likelihood of dataset $d = \{ d_{\rm i}\}$ consisting of individual data $d_i$, $\pi(\Lambda)$ is the prior distribution of $\Lambda$ and $Z_d$ is model evidence.
The total likelihood can be expressed as
\begin{equation}
 \mathcal{L}(d|\Lambda) =  \prod \limits_{i} \mathcal{L}(d_{\rm i}|\Lambda) = \prod \limits_{i} \int \mathcal{L}(d_i | \dot{P}) \pi(\dot{P} | \Lambda) {\rm d}\dot{P},
\end{equation}
where we follow \cite{wph+18} and use a log-uniform conditional prior for $\dot{P}$ (of the MSP population)
 \begin{equation}
	 \label{condition_prior}
    \pi(\dot{P}|\Lambda) = \begin{cases}
        \Big[\dot{P}\ln(\dot{P}_{\rm max}/\dot{P}_{\rm min})\Big]^{-1}, &  \dot{P}_{\rm min} \le \dot{P}\le \dot{P}_{\rm max}(\Lambda) \\
        0, & {\rm otherwise} \\
    \end{cases}
 \end{equation}
with $\dot{P}_{\rm min}$ and $\dot{P}_{\rm max}$ being the minimum and maximum spin-down rate, respectively.
The value of $\dot{P}_{\rm max}$ depends on the model in question. For the spin-up line hypothesis, $\dot{P}_{\rm max}=AP^{4/3}$, while for the alternative hypothesis, $\dot{P}_{\rm max}$ is a constant to be inferred from data.
For a pulsar population, $\dot{P}_{\rm min}$ is conveniently described by a death line.
In principle, we could incorporate the death line into our population model and determine its location.
For simplicity, we opt to use the death line $\dot{P}_{\rm min} = 2.24\times10^{-19} P^{0.98}$ \citep{rr94}, which visually provides a good fit to the pulsar sample considered in this work.
We leave the study of the pulsar death line to a future work.

For the individual likelihood $\mathcal{L}(d_i | \dot{P})$, if the pulsar is isolated and in a globular cluster, its intrinsic spin-down rate $\dot{P}_{\rm int}$ is constrained in a bound, and we assume that $\dot{P}_{\rm int}$ follows a uniform distribution within this bound:  
 \begin{equation}
    \mathcal{L}(d_i | \dot{P}) = \begin{cases}
    \frac{1}{2\sigma_i}, &  \dot{P}_i - \sigma_i \le \dot{P} \le \dot{P}_i + \sigma_i \\
    0, & {\rm otherwise.} \\
    \end{cases}
 \end{equation}
If the pulsar is in the Galactic field or part of a binary in a globular cluster, $\dot{P}_{\rm int}$ follows a Gaussian distribution with a standard deviation given by the measurement uncertainty: 
\begin{equation}
    \mathcal{L}(d_i | \dot{P}) = \frac{1}{\sqrt{2\pi}\sigma_i}\exp\Bigg[ - \frac{(\dot{P}-\dot{P}_i)^2}{2\sigma_i^2}\Bigg].
\end{equation}

We sample the posterior distribution $p(\Lambda | d)$ with the parallel tempering version \citep[\textsc{ptemcee}\footnote{\url{https://pypi.org/project/ptemcee/}},][]{vfm16} of the \textsc{emcee} \citep{fhl+13} software package. The package is capable of calculating model evidence and allows us to perform model selection.
The Bayes factor between two competing models, $\Lambda_1$ and $\Lambda_2$, is
\begin{equation}
	B_{12} = \frac{p(d|\Lambda_1) }{p(d|\Lambda_2)}\, .
\end{equation}	

There are likely to be some selection effects at play in the observed distribution of MSPs in the $P$-$\dot{P}$ diagram.
One such effect that can be modelled is pulsar age.
Assuming that the radio detectability of recycled pulsars evolves slowly since the end of the accretion process, the age-induced bias is roughly proportional to the increase of pulsar age after accretion.
We correct this selection effect by modifying the conditional prior to
\begin{equation}
	\label{eqn:crt_conditional_prior}
	\pi_{\rm crt}(\dot{P}|\Lambda)=\frac{\pi(\dot{P}|\Lambda)}{\tau},
\end{equation}
where $\tau$ is the time spent for the pulsar to evolve from the spin-up line to the present state. For simplicity, we use the characteristic age at the two states to calculate $\tau$, thus $\tau = \tau_{\rm c} - \tau_{{\rm c}, \Lambda}$, where $\tau_{\rm c}=P/(2\dot{P})$, and $\tau_{{\rm c}, \Lambda}$ depends on population hyperparameter $\Lambda$.
Note that in the case of magnetic multipoles, the characteristic age is given by $P/(2l\dot{P})$.
We further assume a constant surface magnetic field strength, i.e.,
\begin{equation}
    P^{2l-1}\dot{P} = {\rm constant}.
\end{equation}
This is a reasonable assumption for MSPs \citep[e.g.,][]{AshleyYuri18}.
Then it can be shown that
\begin{equation}  
	\label{eqn:crt_spinup_line}
    \tau = \frac{P}{2l\dot{P}} - 
    \frac{1}{2l} \mathcal{A}^{-\frac{6l}{4l+3}}\big(\dot{P}P^{2l-1}\big)^{\frac{2l-3}{4l+3}}.
\end{equation}
Here we have added the dimension back to the parameter $\mathcal{A}$ introduced in Equation (\ref{eqn:spin_up_l}) so that it
has a dimension of ${\rm s}^{-2+(2l/3)}$ (only applicable in this equation).

\section{Results} \label{paper_results}
\begin{figure}[htp]
	\centering
	\includegraphics[width=1\columnwidth]{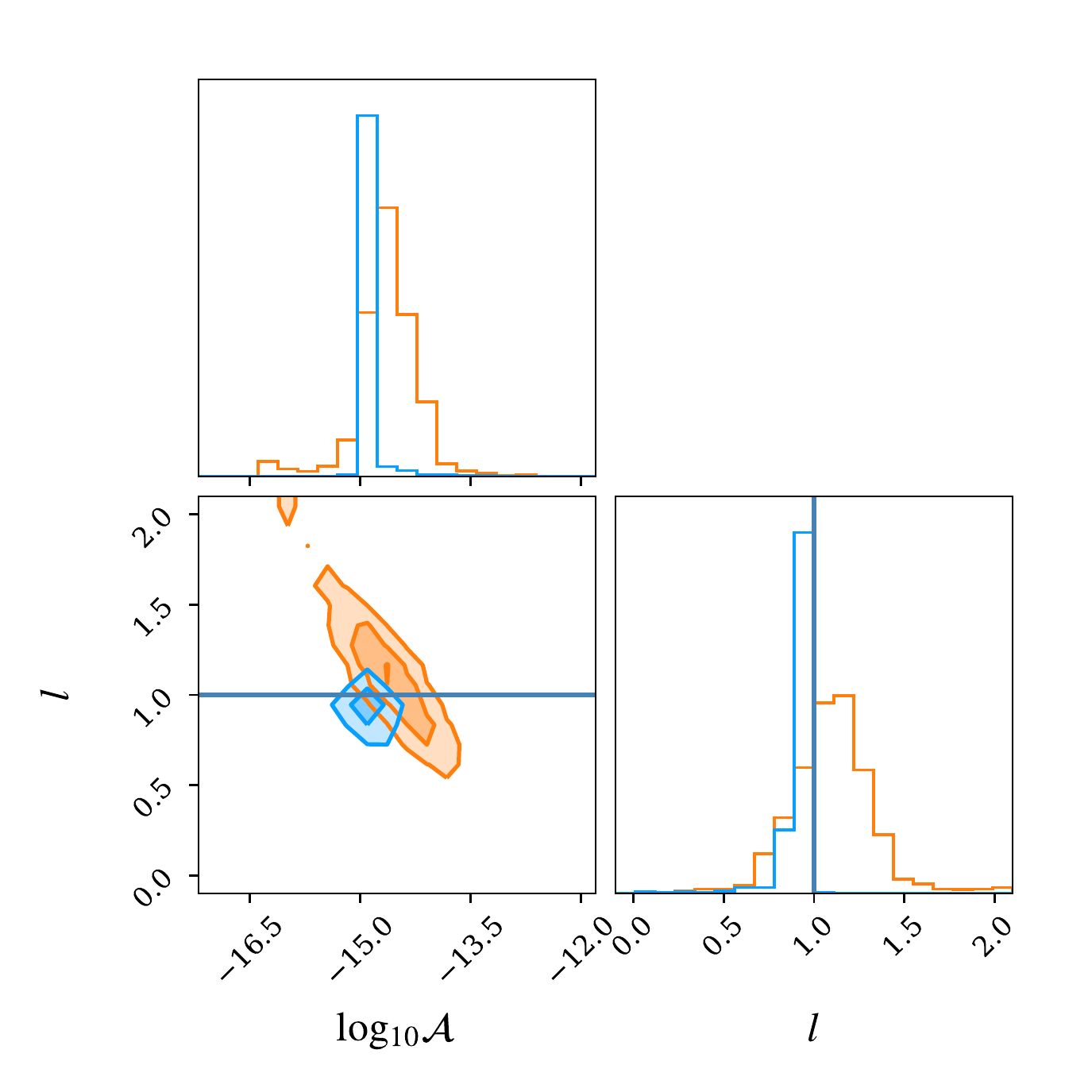}		
	\caption{Joint posterior distribution of the spin-up relation parameter $\mathcal{A}$ and the order of the magnetic multipole $l$. The solid line marks $l=1$ corresponding to pure magnetic dipole. The orange and blue contours are for the case where we include and exclude PSRs J1823$-$3021A and J1824$-$2452A, respectively.}
	\label{fig:Agamma_corner}
\end{figure}

In this section, we present results of Bayesian model selection and parameter estimation.
First, we find that the data strongly favour the presence of a spin-up line, with a natural-log Bayes factor of 20.9 (61.6) by including (excluding) PSRs J1823$-$3021A and J1824$-$2452A. 
In Figure \ref{fig:Agamma_corner}, we show the posterior distribution of $\mathcal{A}$ and $l$ as defined in Equation (\ref{eqn:spin_up_l}); the case of including and excluding PSRs J1823$-$3021A and J1824$-$2452A is plotted in orange and blue, respectively.
We note that in both cases, the posterior distribution of $l$ peaks at $l=1$ (magnetic dipole).
The joint posteriors are broadly consistent in these two cases; the case of excluding those two special pulsars results in much tighter constraints on $\mathcal{A}$ and $l$ because there are more pulsars located near the inferred spin-up line (see Figure \ref{fig:PPdot}).
In both cases, there is no posterior support at $l=3$, which in our analysis is taken as a competing hypothesis for the spin-up line hypothesis.
This is consistent with the large Bayes factors found in our calculations.

The reconstructed spin-up line in the $P$-$\dot{P}$ diagram is also plotted in Figure \ref{fig:PPdot}, with the colored bands indicating the 90\% credible intervals.
In order to compare with the fiducial prediction of a spin-up line, we fix $l=1$ and obtain the 90 percent credible intervals of $A$: $[3.39, 3.76]\times10^{-15}$~s$^{-4/3}$ when PSRs J1823$-$3021A and J1824$-$2452A are included and $[8.87, 9.34]\times10^{-16}$~s$^{-4/3}$ when both pulsars are excluded.
In comparison, the classic spin-up line predicts $A=(1.1 \pm 0.5)\times10^{-15}$~s$^{-4/3}$.

\begin{figure*}
		\includegraphics[width=\textwidth]{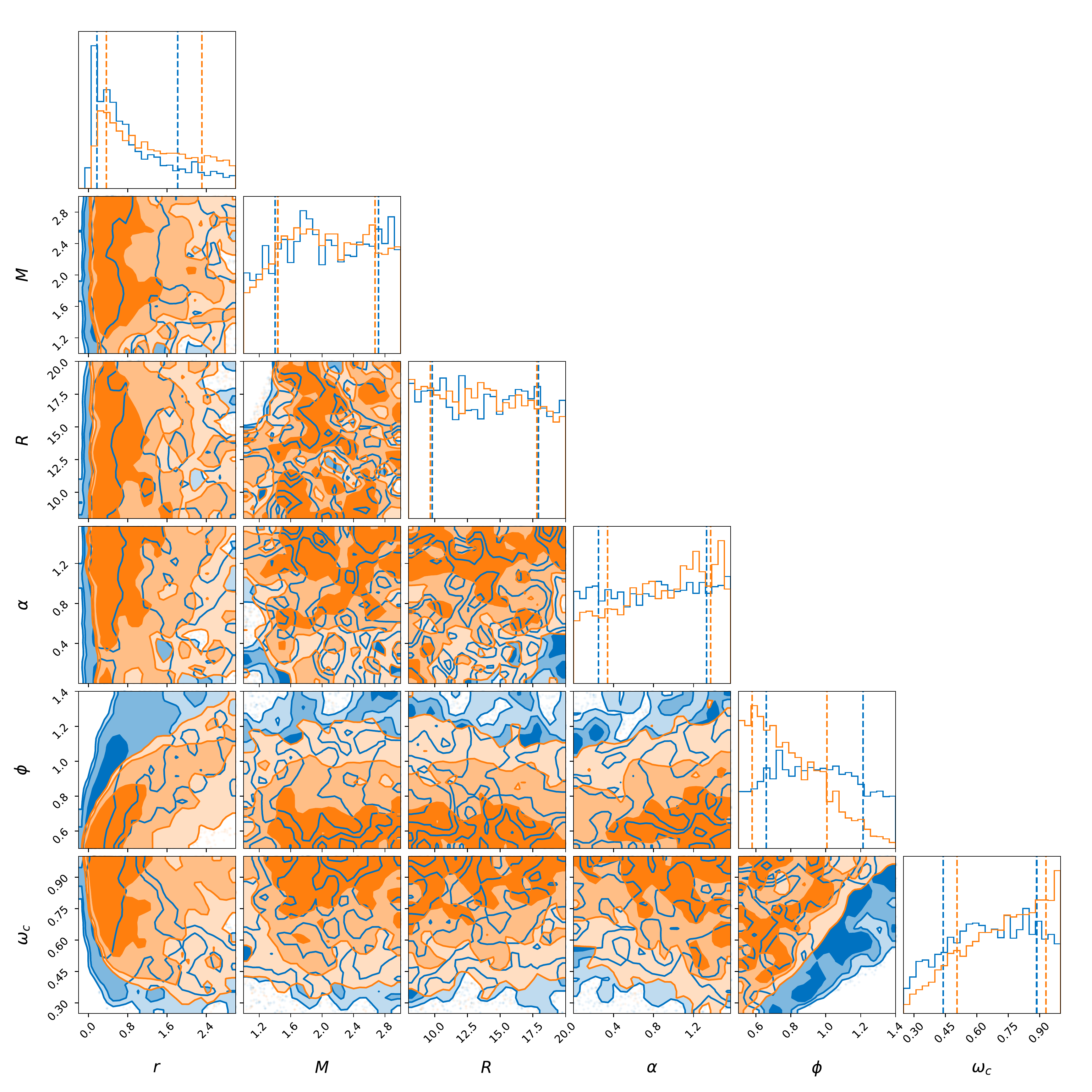}
		\caption{Posterior distributions of the accretion parameters defined in Equation (\ref{eqn:spin-up-param}). Posteriors obtained using all the $P$-$\dot{P}$ data are in orange, while those obtained by excluding PSRs J1823$-$3021A and J1824$-$2452A are in blue. The intervals on the 1D posterior distributions are at $1\sigma$.}
		\label{fig:accretion_corner}
\end{figure*}

\begin{table*}[htp] 
	\caption{The prior ranges and posteriors (median and $1\sigma$ intervals) of accretion parameters: $r$ is the accretion rate in units of Eddington accretion rate, $M$ and $R$ is pulsar mass and radius, $\alpha$ is magnetic inclination angle, $\phi$ is a parameter describing the size of magnetosphere and $\omega_{\rm c}$ is the critical fastness parameter. Both $\phi$ and $\omega_{\rm c}$ are dimensionless. The posteriors are given in two cases, where case $a$ uses all the $P$-$\dot{P}$ data and case $b$ excludes PSRs J1823$-$3021A and J1824$-$2452A.}
	\begin{center}
		\begin{tabular}{lcccccc}
		\hline\hline		
		 & \multicolumn{6}{c}{Accretion parameters defined in Equation (\ref{eqn:spin-up-param})} \\
		\cline{2-7}
		   & $r$ & $M$ ($M_\odot$) & $R$ (km) & $\alpha$ (rad) & $\phi$ & $\omega_{\rm c}$ \\
		 \hline
		 prior & [0.01, 3] & [1, 3] & [8, 20] & [0, $\pi/2$] & [0.5, 1.4] & [0.25, 1] \\
		 posterior (case $a$) &
		$ 1.13^{+1.18}_{-0.76}$ &
		$ 2.04^{+0.63}_{-0.61}$ &
		$ 13.70^{+4.10}_{-4.03}$ &
		$ 0.91^{+0.46}_{-0.57}$ &
		$ 0.76^{+0.25}_{-0.18}$ &
		$ 0.75^{+0.18}_{-0.24}$ \\
        posterior (case $b$) &
        $0.65^{+1.17}_{-0.48}$ &
        $2.03^{+0.69}_{-0.63}$ &
        $13.92^{+4.00}_{-4.11}$ &
        $0.82^{+0.51}_{-0.57}$ &
        $0.93^{+0.28}_{-0.27}$ &
        $0.67^{+0.22}_{-0.23}$ \\
		\hline
		\hline
		\end{tabular}
		\label{tab:spinup_param}
	\end{center}
\end{table*}

Next, we fix $l=1$ (i.e., assuming pure magnetic dipole) and derive constraints on accretion parameters defined in Equation (\ref{eqn:spin-up-param}).
Figure \ref{fig:accretion_corner} shows the posterior distributions, with orange and blue colors representing the case where we include and exclude PSRs J1823$-$3021A and J1824$-$2452A in the analysis, respectively.
The $1\sigma$ intervals of marginalized 1D posterior distributions are listed in Table \ref{tab:spinup_param}.
We see from Figure~\ref{fig:accretion_corner} that the inclusion of PSRs J1823$-$3021A and J1824$-$2452A results in considerable deviations in the posterior distributions of $\phi$ and $\omega_{\rm c}$ from their prior distributions, which are assumed to flat.
This is because a larger $A$ is required in this case and there is strong dependence of $A$ on those two parameters.
In \citet{tlk12}, it was shown that small changes in $\phi$ and $\omega_{\rm c}$ could shift the spin-up line by up to an order of magnitude.
On the other hand, the inclusion of PSRs J1823$-$3021A and J1824$-$2452A makes a negligible impact on the estimation of $M$, $R$ and $\alpha$, because of a weak dependence of the spin-up line on $M, R$ and $\alpha$; see Equations~(\ref{eqn:spin-up-param}) and~(\ref{eqn:inertia}).

\section{Discussion and Conclusions}
\label{sec:conclu}

Our analysis shows that PSRs J1823$-$3021A and J1824$-$2452A, having unusually high spin-down rate $\dot{P}\gtrsim 10^{-18}$, can be accommodated in a single population with other MSPs. 
PSRs J1823$-$3021A and J1824$-$2452A are in globular clusters NGC~6624 and M28, respectively, and both have energetic $\gamma$-ray emission, whose efficiency implies that both $\dot{P}_{\rm obs}$ are largely intrinsic \citep{faa+11, jgk+13}. 
Their high spin-down rates and strong $\gamma$-ray emission also suggest small characteristic ages.
This is consistent with the result from our analysis in the sense that both pulsars are located near the spin-up line, which is expected if both pulsars completed the recycling process not long before, possibly with a near-Eddington accretion rate.
Fast MSPs in the region of $P \lesssim \unit[5]{ms}$ and $\dot{P} \gtrsim 10^{-19}$~s~s$^{-1}$ are likely to be discovered in current and future pulsar surveys. The lack of such MSPs known so far could be due to a selection effect as they spend a smaller part of their lifetimes in this parameter space.

Sub-millisecond pulsars have long been suggested to exist \citep[e.g.,][]{DXQ+09} although none has been discovered so far.
Recently, \citet{wph+18} showed that there is likely to be a constraint set by gravitational-wave spin down due to a minimum ellipticity (shown as dashed line in Figure \ref{fig:PPdot}).
Combining this with the spin-up line found in this work implies a minimum spin period of $\sim \unit[0.6]{ms}$.

In this work, we report observational evidence for a spin-up line in the pulsar $P$-$\dot{P}$ diagram and demonstrate its usefulness in constraining accretion physics.
We assume that $P$ and $\dot{P}$ of MSPs follows a log-uniform distribution, which is a reasonable model for a relatively small set of observations.
A more sophisticated model might involve: 1) supposing a fraction of MSPs reach the equilibrium spin periods at the end of accretion and thus start their life as a recycled pulsar from the spin-up line, and the remaining fraction might end up in a non-equilibrium spin-up line \citep[e.g.,][]{acw99}; 2) MSPs are evolved to their current locations in the $P$-$\dot{P}$ diagram due to intrinsic spin-down.
Therefore, the observed distribution of MSPs in the $P$-$\dot{P}$ diagram will allow one not only to probe accretion physics, but also to constrain pulsar spin evolution and the minimum spin period of neutron stars.

\begin{acknowledgments}
We thank the referee for very useful comments that helped improve our work.
We also thank Dr Zu-Cheng Chen for useful discussion.
This work is supported by a research start-up grant provided by Beijing Normal University at Zhuhai.

\end{acknowledgments}

%



\software{	Jupyter \citep{krp+16}, Numpy \citep{hm20}, Matplotlib \citep{hun07}, Astropy \citep{art+13, aps+18}, ptemcee \citep{vfm16}.
          }



%
%


\bibliography{spinup}{}
\bibliographystyle{aasjournal}

\appendix

\begin{figure}
    \centering
    \includegraphics[width=0.46\columnwidth]{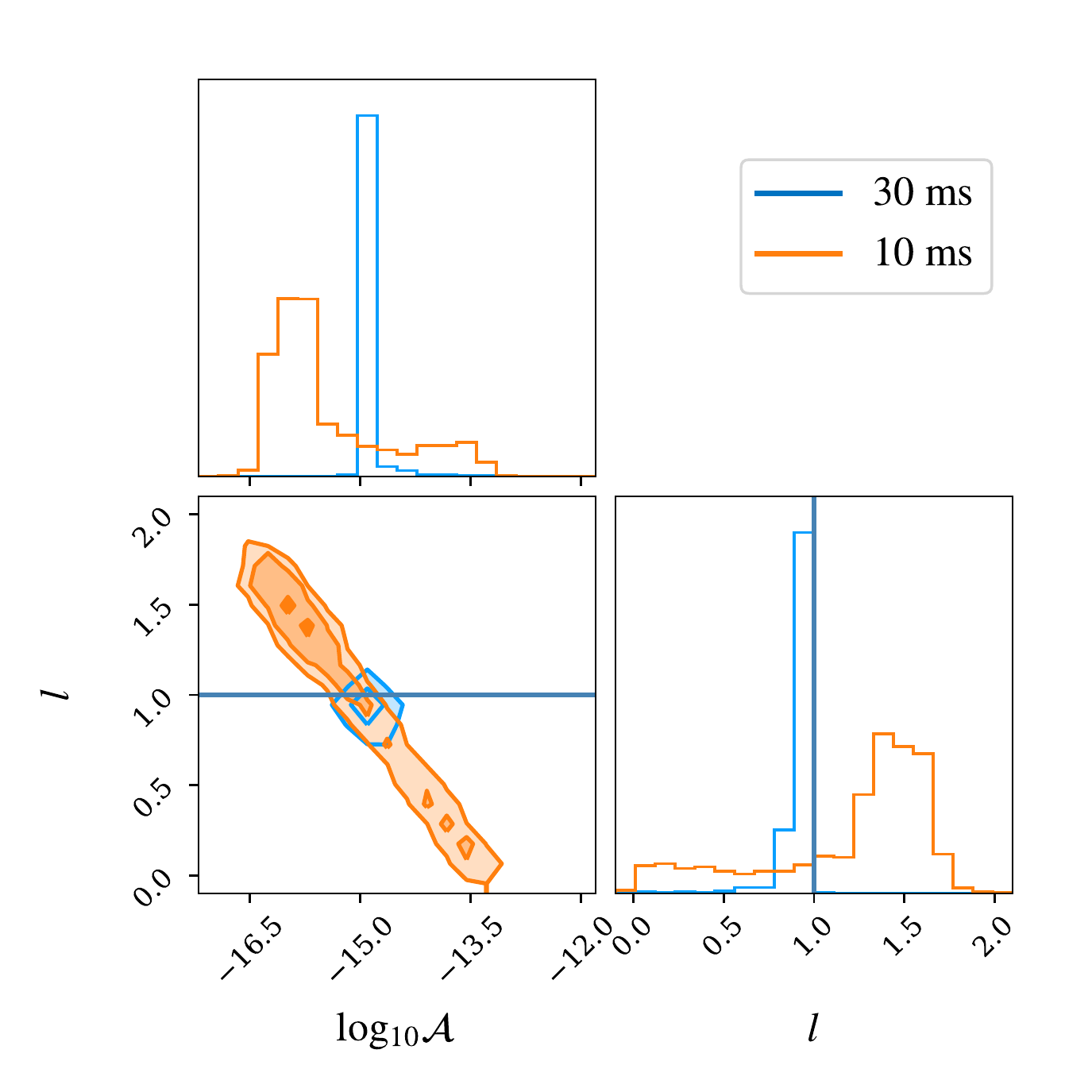}
    \caption{A comparison of the joint posterior distribution of $\mathcal{A}$ and $l$ for different MSP spin period cuts. The blue contour is taken from Figure~\ref{fig:Agamma_corner} ($P<30$~ms), while the orange one is for a smaller sample with 109 pulsars of $P<10$~ms. In both cases, PSRs~J1823$-$3021A and J1824$-$2452A are excluded.}
    \label{fig:10ms_30ms_Al}
\end{figure}

The term of a millisecond pulsar usually refers to a recycled pulsar with a spin period of $P \lesssim \unit[30]{ms}$.
Here we discuss the inference results if we apply a lower spin period cut at $\unit[10]{ms}$.
In Figure \ref{fig:10ms_30ms_Al} we compare the joint posterior distributions of $\mathcal{A}$ and $l$ for the two cases with a period cut at $\unit[30]{ms}$ (blue) and $\unit[10]{ms}$ (orange); in both cases PSRs~J1823$-$3021A and J1824$-$2452A are excluded in the analysis.
One can see that using a larger sample of pulsars leads to tighter constraints on the spin-up line parameters. In particular, there are a couple more pulsars that are located nearly on the inferred spin-up line (see Figure \ref{fig:PPdot}).
We also find that even if we use only pulsars with $P< \unit[10]{ms}$, the evidence for a spin-up line is very strong, with a natural-log Bayes factor of 11.5.
Finally, in cases where we include PSRs~J1823$-$3021A and J1824$-$2452A in the analysis, we find negligible difference in posterior distributions between $P< \unit[10]{ms}$ and $P< \unit[30]{ms}$. This is unsurprising because the location of the spin-up line is almost exclusively determined by those two pulsars as can be seen in Figure \ref{fig:PPdot}.


In this work, we find that the observational evidence for a spin-up line is dominated by Galactic-field MSPs.
Our MSP sample also includes ten MSPs in globular clusters (plotted as green or red filled circles in Figure \ref{fig:PPdot}; see also Table~\ref{tab:GC_iso_MSP}).
They seem to be consistent with the broad MSP population.
Here we comment on the relatively small uncertainties of their intrinsic spin-down rates $\dot{P}_{\rm int}$.
First, we discuss four isolated cluster MSPs.
For PSRs~J1823$-$3021A and J1824$-$2452A, their measured high $\gamma$-ray luminosity implies that their observed spin-down rates are largely intrinsic.
This is consistent with our derived ranges of $\dot{P}_{\rm int}$ for the two pulsars.
A similar scenario applies to PSR J1910$-$5959D, which has the third highest $\dot{P}_{\rm obs}$ among the cluster MSPs considered here.
In \citet{AmicoJ19105959D}, various scalings between the X-ray luminosity and spin-down power for MSPs were applied to PSR J1910$-$5959D to estimate $\dot{P}_{\rm int}$; they found a lower bound at $6.8\times 10^{-19}$ (see their fig. 2), which is comparable to our lower bound of $7.8\times 10^{-19}$. Another isolated cluster pulsar, J1518+0204A, also has a relatively small error of $\dot{P}_{\rm int}$ as inferred from its host cluster potential. This is expected because the pulsar is located far from the centre of its parent cluster; the core size of the cluster is $0\farcm44$ while the pulsar has an angular offset of $0\farcm50$ from the cluster centre\footnote{\url{http://www.naic.edu/~pfreire/GCpsr.html}}. 

Among six binary MSPs in globular clusters, three are redback or black widow systems and do not have measured $\dot{P}_{\rm b}$. 
We thus constrain their $\dot{P}_{\rm int}$ in the same way as isolated cluster pulsars.
For PSR J1740$-$5340A, we find that its $\dot{P}_{\rm int}$ can be constrained to be between $1.4$ and $2.0 \times 10^{-19}$; noting that $\dot{P}_{\rm obs}=1.68 \times 10^{-19}$, the correction due to cluster potential is small as the pulsar is far away from its cluster centre; the position offset is $0\farcm55$ whereas the cluster core radius is only $0\farcm05$ \citep{PSRJ1740-5340A}.
Our corrections for cluster potential for PSRs J1701$-$3006E and J1701$-$3006F (both located in M62) are consistent with those reported in the original publication \citep{lfr+12};
we find upper bounds at $5.0$ and $3.7 \times 10^{-19}$ for PSRs E and F, respectively; the corresponding bounds are $5.56$ and $4.14 \times 10^{-19}$ in \citet{lfr+12}. Finally, the remaining three pulsars in Table~\ref{tab:GC_iso_MSP} are in binary systems and have measured $\dot{P}_{\rm b}$ \citep{frk+17}. 

\begin{table}[htp] 
	\caption{Properties of MSPs in globular clusters, including the spin period $P$, the observed spin-down rate $\dot{P}_{\rm obs}$ (in unit of $10^{-19}$), the possible range of intrinsic spin-down rate $\dot{P}_{\rm int}$ (in unit of $10^{-19}$), the type of error used for performing population inference (``uniform" indicates a uniform distribution within the given range; the range for a Gaussian error is given for the mean $\pm 1 \sigma$), the pulsar type, and the name of host cluster. The pulsars labelled with $^\dagger$ and $^\ast$ are redback and black widow systems, respectively.}
	\begin{center}
	\begin{tabular}{lcccccc}
		\hline
		\hline
	Pulsar name  & $P$ (ms) & $\dot{P}_{\rm obs}$ & Range of $\dot{P}_{\rm int}$ & Error & Type & Cluster \\
	\hline
J1823$-$3021A & 5.44 & $33.8$ & $[31.0,37.0]$ & Uniform & isolated & NGC 6624 \\
J1824$-$2452A & 3.05 & $16.2$ & $[15.0,17.0]$ & Uniform & isolated & M28 \\
J1910$-$5959D & 9.04 & $9.64$ & $[7.8,11.4]$ & Uniform & isolated & NGC 6752\\
J1518+0204A & 5.55 & 0.41 & $[0.23, 0.59]$ & Uniform & isolated & M5 \\ 

J1740$-$5340A$^\dagger$ & 3.65 & $1.68$ & $[1.4,2.0]$ & Uniform & binary & NGC 6397 \\
J1701$-$3006E$^\ast$ & 3.23 & $3.10$ & $[1.2,5.0]$ & Uniform & binary & M62 \\
J1701$-$3006F$^\ast$ & 2.30 & $2.22$ & $[0.7,3.7]$ & Uniform & binary & M62 \\

J0024$-$7204T & 7.59 & 2.94 & $[0.10, 1.88]$ & Gaussian & binary & 47 Tuc \\
J0024$-$7205E & 3.54 & 0.99 & $[0.08, 0.16]$ & Gaussian & binary & 47 Tuc \\
J0024$-$7203U & 4.34 & 0.95 & $[0.13, 0.24]$ & Gaussian & binary & 47 Tuc \\
	\hline
	\hline
	\end{tabular}
	\label{tab:GC_iso_MSP}
	\end{center}
\end{table}


\clearpage\end{CJK*}
\end{document}